\documentclass[aps,prb,twocolumn,superscriptaddress,showpacs]{revtex4-1}

\usepackage{graphicx}

\begin{document}

\title{Framework Flexibility and the Negative Thermal Expansion Mechanism of Copper(I) Oxide, Cu$_2$O}
\author{Leila H. N. Rimmer}
\affiliation{Department of Earth Sciences, University of Cambridge, Downing Street, Cambridge CB2 3EQ, U.K.}

\author{Martin T. Dove}
\email{martin.dove@qmul.ac.uk}
\affiliation{Department of Earth Sciences, University of Cambridge, Downing Street, Cambridge CB2 3EQ, U.K.}
\affiliation{Centre for Condensed Matter and Materials Physics, School of Physics and Astronomy, Queen Mary University of London, Mile End Road, London E1 4NS, U.K.}
\affiliation{Materials Research Institute, Queen Mary University of London, Mile End Road, London E1 4NS, U.K.}

\author{Bj\"orn Winkler}
\affiliation{Geowissenschaften, Goethe-UniversitŠt, Altenhoeferallee 1, D-60438 Frankfurt a.M., Germany.}

\author{Dan J. Wilson}
\affiliation{Geowissenschaften, Goethe-UniversitŠt, Altenhoeferallee 1, D-60438 Frankfurt a.M., Germany.}

\author{Keith Refson}
\affiliation{Science and Technology Facilities Council, Rutherford Appleton Laboratory, Harwell Science and Innovation Campus, Didcot OX11 0QX, U.K.}

\author{Andrew L. Goodwin}
\affiliation{Inorganic Chemistry Laboratory, University of Oxford, South Parks Road, Oxford OX1 3QR, U.K.}

\date{\today}

\begin{abstract}
The negative thermal expansion (NTE) mechanism in Cu$_2$O has been characterised via mapping of different Cu$_2$O structural flexibility models onto phonons obtained using \emph{ab-initio} lattice dynamics. Low frequency acoustic modes that are responsible for the NTE in this material correspond to vibrations of rigid O--Cu--O rods. There is also some small contribution from higher frequency optic modes that correspond to rotations of rigid and near-rigid OCu$_4$ tetrahedra as well as of near-rigid O--Cu--O rods. The primary NTE mode also drives a ferroelastic phase transition at high pressure; our calculations predict this to be to an orthorhombic structure with space group $Pnnn$.
\end{abstract}


\maketitle

\section{Introduction}
\label{Introduction}

Negative thermal expansion (NTE) is a property of significant interest to those working in the field of materials design. Not only does it offer the prospect of structures that remain undistorted over significant temperature ranges\cite{Romao_etal_2013_NTEChapter,Lind_2012_Mater_NTEReview,Takenaka_NTEApplicationsReview,Miller_etal_2009_JMatSci_NTEReview,Goodwin_2008_NatNano_NTE} but, in addition, other unusual phenomena (such as softening under pressure or enhanced gas adsorption and filtration) are often linked to the microscopic origins of NTE.\cite{Fang_Dove_2013_PRB_PISZeolites,Hammonds_etal_1998_JPhysChemB_Zeolites} Understanding these microscopic driving mechanisms in NTE materials is therefore of fundamental importance.

\begin{figure}[hbt]
	\includegraphics[width=0.45\textwidth]{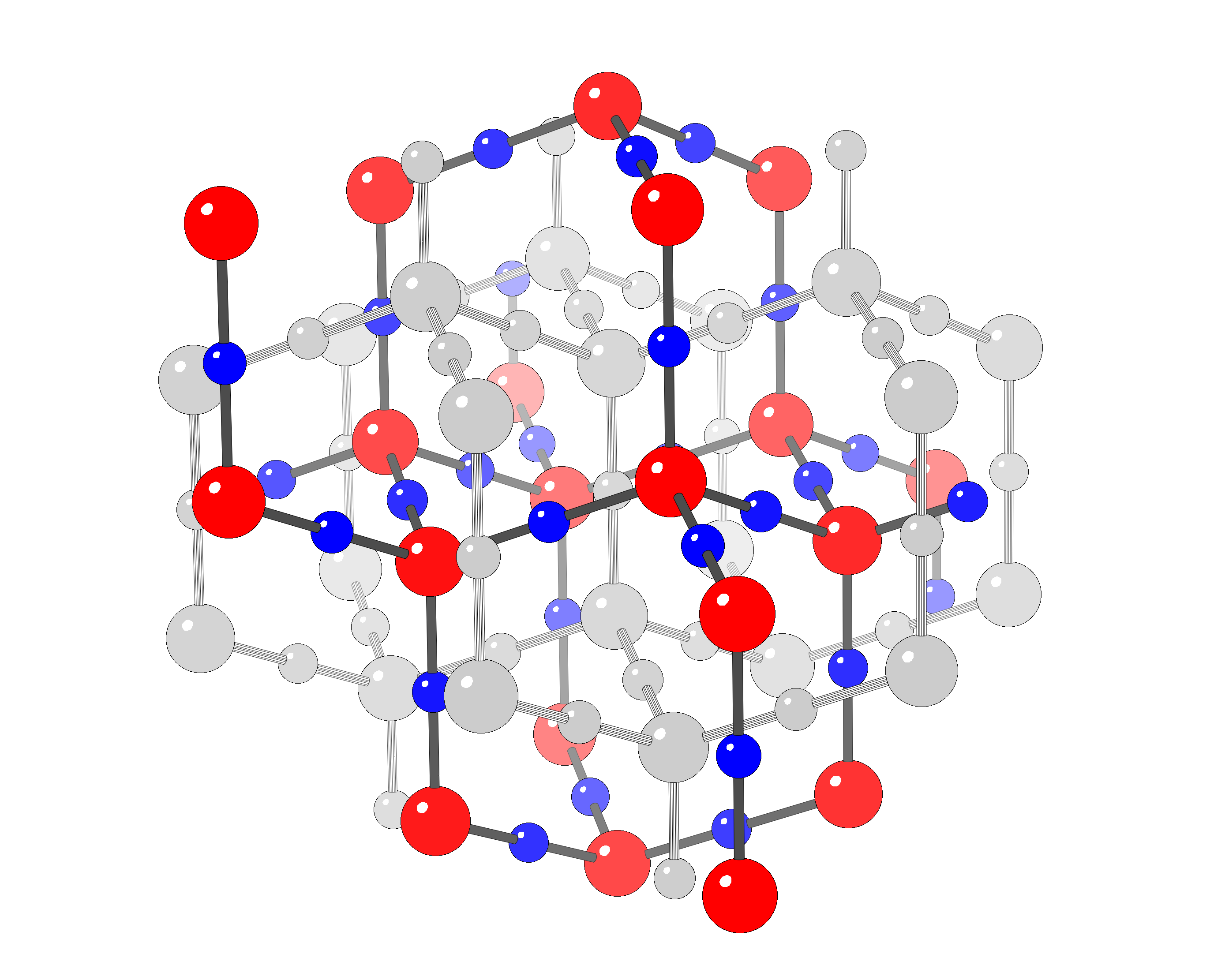}
	\caption{\label{cu2ointro}The Cu$_2$O crystal framework. The structure is shown in ball and stick representation: red atoms are O; blue atoms are Cu. One of the two sublattices is shaded in pale grey for clarity.}
\end{figure}

Copper oxide, Cu$_2$O, has NTE at low temperatures. Its coefficient of thermal expansion has been variously measured as having a minimum value between $-$8.0 and $-$2.95~MK$^{-1}$ at temperatures between 80 and 100~K, before increasing to become weak positive thermal expansion above 200~K.\cite{White_1978_JPhysC_Cu2ONTE,Schafer_Kirfel_2002_ApplPhysA_Cu2ONeutrons,Dapiaggi_etal_2003_XRD-EXAFSNTEinCu2O,Chapman_Chupas_2009_ChemMater_NTECuprites}

Cu$_2$O is also one of the simpler NTE-exhibiting frameworks,\cite{Fornasini_etal_2009_NTELocalDynamics} its structure consisting of two interpenetrating diamondoid cristobalite lattices---see Figure \ref{cu2ointro}. Despite its structural simplicity compared to other NTE frameworks, the atomic-scale origin of the NTE in Cu$_2$O is not fully understood.

Inelastic neutron scattering,\cite{Bohnen_etal_2009_Cu2ONTE} Raman scattering and luminescence\cite{Reimann_Syassen_1989_PhysRevB_Cu2ORamanPhotoluminescence} as well as \emph{ab-initio} calculations\cite{Bohnen_etal_2009_Cu2ONTE,Gupta_etal_2013_ArXiv_AbInitNTEPhonons} have identified specific phonon modes responsible for driving NTE in Cu$_2$O. All such modes exist below 5~THz, with the largest contribution coming from transverse acoustic modes in the $\Gamma$--X--M region of reciprocal space\cite{Bohnen_etal_2009_Cu2ONTE,Gupta_etal_2013_ArXiv_AbInitNTEPhonons} (in this paper we use the notation of Bradley and Cracknell\cite{Bradley_Cracknell_1972_CrystalBZSymmetryBook} to label the high-symmetry points of the reciprocal cell). Optic modes that contribute weakly to both NTE and positive thermal expansion (PTE) also exist in this energy range.\cite{Reimann_Syassen_1989_PhysRevB_Cu2ORamanPhotoluminescence,Bohnen_etal_2009_Cu2ONTE,Gupta_etal_2013_ArXiv_AbInitNTEPhonons}

\begin{figure}[tb]
	\includegraphics[width=0.45\textwidth]{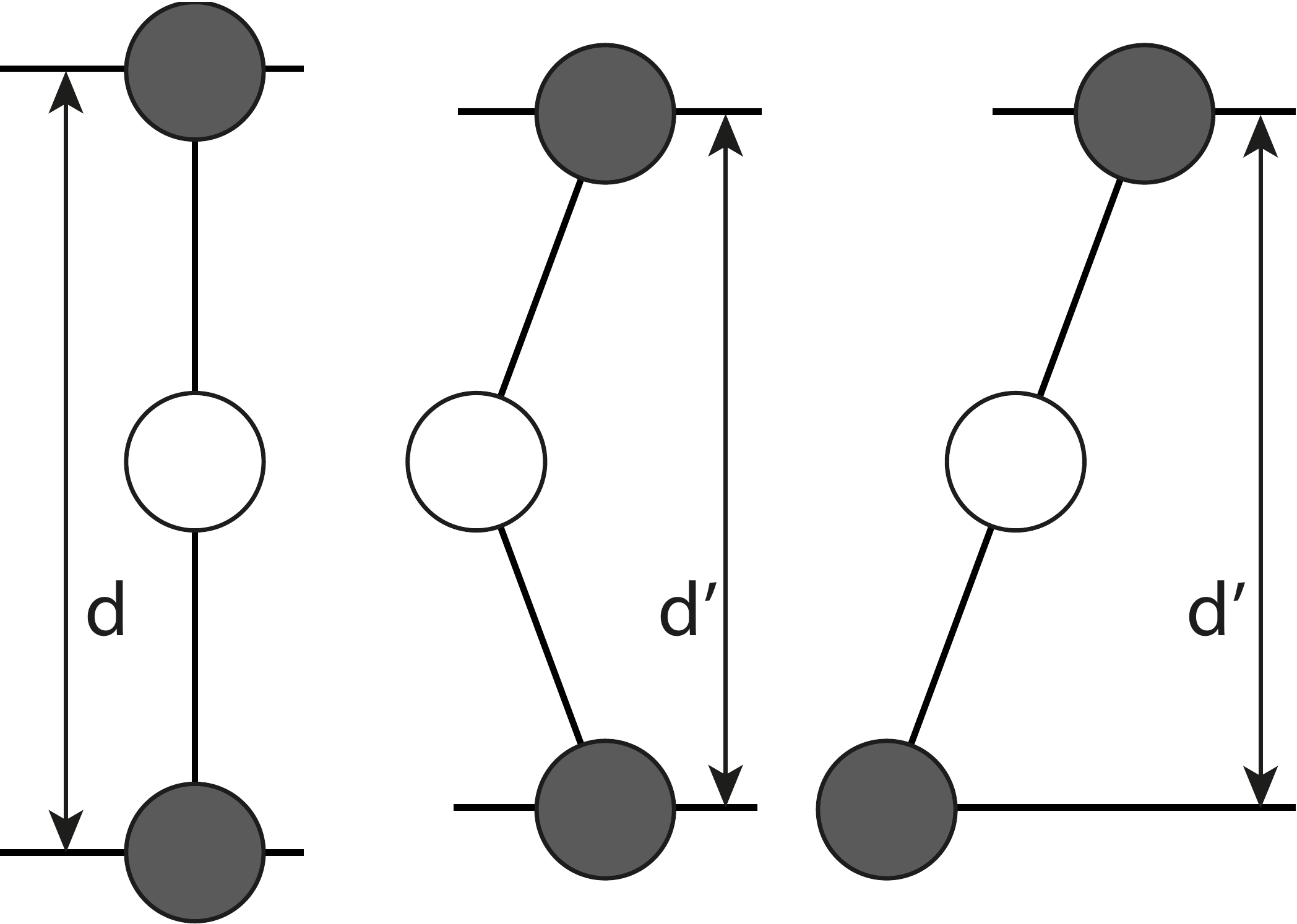}
	\caption{\label{tensioneffect}Illustration of two transverse vibrations that can contribute to  NTE. Both lead to negligible stretching of nearest-neighbour bonds, but result in reduction of the inter-atomic or inter-planar distances (in this case reducing distance $d$ to $d'$). If enough of these vibrations exist with large enough amplitude in a material, they can drive macroscopic NTE.}
\end{figure}

A common mechanism for NTE suggested for framework materials is the `tension effect',\cite{Barrera_etal_2005_JPhysCondMat_NTEReview} in which transverse vibrations of bridging atoms in a three-atom connection bring the two end atoms towards each other due to the relatively high energy cost of stretching interatomic bonds. \cite{Romao_etal_2013_NTEChapter,Miller_etal_2009_JMatSci_NTEReview,Sleight_1998_AnnuRevMatSci_IsotropicNTE}  However, there is a similar mechanism in which a group of atoms rotate as a whole and bring inwards the planes of atoms connected to their ends. Both mechanisms are shown in Figure \ref{tensioneffect}. A similar mechanism is found to be important in the related NTE material Zn(CN)$_2$.\cite{Fang_etal_2013_PRB_NTEZnCN2} Both types of motions will be correlated and described in terms of phonon normal modes. For zero wave vector, the traditional tension mechanism will be described by optic phonons, but the rotational motion will more likely be part of a transverse acoustic phonon that generates a shear deformation of the structure.

All materials undergo some drive towards PTE due to longitudinal bond expansion arising from the anharmonicity of interatomic potentials.\cite{Dove_IntroductiontoLatticeDynamics} However, if a tension-effect mode has low enough energy and thus large enough amplitude (given that a phonon's amplitude is inversely proportional to the square of its frequency\cite{Dove_IntroductiontoLatticeDynamics}) and if it comprises sufficiently large a fraction of the full phonon spectrum, then the PTE is outweighed and the net outcome is macroscopic NTE.\cite{Miller_etal_2009_JMatSci_NTEReview}

Parametrised potential calculations\cite{Mittal_etal_2007_PRB_Cu2ONeutronsandCalculations} have identified an optic phonon that dominates the low energy density of states. Analysis of its eigenvectors at the $\Gamma$ point\cite{Mittal_etal_2007_PRB_Cu2ONeutronsandCalculations,Sanson_2011_AbInitCu2OAg2O} show this mode to consist of rotations of undeformed OCu$_4$ tetrahedra. This is a type of tension effect known as a \emph{rigid unit mode}\cite{Hammonds_etal_1996_AmMineral_QRUMPaper} (RUM), an established NTE mechanism in some silicate minerals.\cite{Heine_etal_199_JAmCerSoc_NTETheory,Welche_etal_1998_PhysChemMater_NTEBetaQuartz} The same mechanism has also been proposed as a result of evidence from neutron powder diffraction\cite{Schafer_Kirfel_2002_ApplPhysA_Cu2ONeutrons} where anisotropy of Cu thermal displacement parameters points to significant Cu motion transverse to the $\langle 111\rangle$ directions.

Meanwhile EXAFS work,\cite{Chapman_Chupas_2009_ChemMater_NTECuprites,Fornasini_etal_2009_NTELocalDynamics,Forasini_etal_2008_EXAFSandNTE,Fornasini_etal_LocalDynamicsNTE,Fornasini_etal_2006_LocalBehaviourNTE,Sanson_etal_2006_PRB_Cu2OAg2ONTE} as well as X-ray\cite{Chapman_Chupas_2009_ChemMater_NTECuprites} and neutron\cite{Dapiaggi_etal_2008_JPhysChemSolid_Cu2ONTEPDFs} PDF studies, have led to proposals of an alternative tension effect mechanism involving significant OCu$_4$ tetrahedral deformations, either in addition to the RUM \cite{Chapman_Chupas_2009_ChemMater_NTECuprites} or replacing it.\cite{Sanson_2009_SolStatSci_TensionLocalDynamicsCuprites,Sanson_etal_2006_PRB_Cu2OAg2ONTE} This is due to observations of both expansion and contraction in average Cu$\cdots$Cu distance as a function of increasing temperature, interpreted as deformations of the OCu$_4$ tetrahedral units.

However, no proposed mechanism has been tested against the known NTE phonons in Cu$_2$O. It is not enough to determine that a type of large amplitude structural deformation (and therefore possible tension effect mechanism) takes place in the system; in order to understand the driving force behind its NTE, we \emph{must} find the specific type of tension effect (if any) that corresponds to its  NTE-driving modes.

Our approach to this problem is to map phonons from different models of Cu$_2$O structural flexibility onto those from high quality \emph{ab-initio} calculations. By observing the degree to which each flexibility model is able to describe the real system's NTE modes we can thus determine the type of framework flexibility, and the particular manifestation of the tension effect, that drives the NTE in Cu$_2$O.

\section{Methods}

\subsection{\emph{Ab-initio} lattice dynamics}
\label{Abinitiolatticedynamics}

\begin{table}[bt]
\caption{\label{CompareWithOtherCalcs} Cu$_2$O equilibrium cell parameters, as obtained through calculation and experiment. Our value of $a=4.358$~\AA\ is a $\sim$2\% overestimation of the experimentally-derived values, a typical result for GGA calculations.
The symbol \dag\ indicates that the measurements were performed at 300~K.
The symbol \ddag\ indicates that the measurements were performed at 15~K.}
\begin{ruledtabular}
\begin{tabular}{l c c}
Source & a$_0$~/~\AA \\
\hline
\hline
This work (PBE+pseudopotentials) & 4.358 \\
PBE+pseudopotentials\cite{Cortona_Mebarki_2011_Cu2OHighP} & 4.359 \\
LDA+pesudopotentials\cite{Cortona_Mebarki_2011_Cu2OHighP} & 4.221 \\
PBE+mixed-basis pseudopotentials\cite{Bohnen_etal_2009_Cu2ONTE} & 4.30 \\
All-electron calculation\cite{Bohnen_etal_2009_Cu2ONTE} & 4.32 \\
X-ray powder diffraction\cite{Sanson_etal_2006_PRB_Cu2OAg2ONTE}\dag & 4.27014(7) \\
Neutron powder diffraction\cite{Schafer_Kirfel_2002_ApplPhysA_Cu2ONeutrons}\ddag & 4.2763(2) \\
\end{tabular}
\end{ruledtabular}
\end{table}

Density functional theory\cite{Hohenberg_Kohn_1964_PhysRev_OriginalDFTPaper1,Kohn_Sham_1965_PhysRev_DFTKohnShamEquations} simulations were performed using the plane-wave pseudopotential method\cite{Payne_etal_1992_RevModPhys_PlaneWaveDFTPaper} as implemented in CASTEP.\cite{Clark_etal_2005_ZKrist_CASTEP} The GGA-PBE functional\cite{Perdew_etal_1996_PRL_PBEFunctional,Perdew_etal_1997_PRL_PBEFunctionalErrata} was used for all calculations.

Troullier-Martins norm-conserving pseudopotentials describing Cu and O were obtained from the ABINIT FHI database.\cite{Fuchs_etal_1999_CompPhysCommun_FHIPseudopotentials,Abinit_PseudopotentialLibrary} The density mixing procedure, with 20 densities stored in its history, was implemented for SCF minimisation. A plane-wave cutoff energy of 1100~eV and corresponding FFT grid were used to define the basis for the electronic orbitals. To minimise violation of the acoustic sum rule arising from high-frequency components of the GGA XC functional, a grid 2.5 times denser in each direction was used to represent the electron density and potentials. Electronic Brillouin-zone integrals were evaluated using a $12\times12\times12$ Monkhorst-Pack grid. In addition, Cu$_2$O was explicitly defined as a zero-temperature insulator by keeping band occupancies fixed.

\begin{table*}[tb]
\caption{\label{ComparePhononsWithOtherCalcs}
Cu$_2$O phonon frequencies at the $\Gamma$ point for the equilibrium Cu$_2$O cell volume, as obtained through calculation and experiment. Frequencies are in units of cm$^{-1}$. Calculated frequencies are within a few percent of values obtained via experiment, a typical result for DFT. 
The symbol \dag\ indicates data obtained via Reference \onlinecite{Sanson_2011_AbInitCu2OAg2O}.}
\begin{ruledtabular}
\begin{tabular}{l c c c c c c c c}
Study & T$_{2u}$ & E$_u$ & T$_{1u}$ (TO) & T$_{1u}$ (LO) & B$_u$ & F$_{2g}$ & F$_{1u}$ (TO) & F$_{1u}$ (LO) \\
\hline
\hline
This work (DFPT calculation) & 64 & 77 & 139 & 140 & 337 & 489 & 607 & 629 \\
DFPT calculation\cite{Bohnen_etal_2009_Cu2ONTE}\dag & 72 & 86 & 147 & 148 & 337 & 496 & 609 & 629 \\
All-electron finite displacement calculation\cite{Sanson_2011_AbInitCu2OAg2O} & 67 & 119 & 142 & 146 & 350 & 515 & 635 & 654 \\
Raman+luminescence\cite{Reimann_Syassen_1989_PhysRevB_Cu2ORamanPhotoluminescence}  & 86 & 110 & 152 & 152 & 350 & 515 & 633 & 662 \\
\end{tabular}
\end{ruledtabular}
\end{table*}

A geometry optimisation was performed at 0~GPa such that residual stresses on the cell were within 0.02~GPa. Since the atomic coordinates of Cu$_2$O are fixed by its $Pn\bar{3}m$ symmetry, only the cell parameter required optimisation. Table \ref{CompareWithOtherCalcs} shows the converged result of $a=4.358$~\AA. This is a 2\% overestimation of the experimental value, as is typical for GGA-PBE.

Phonons were calculated for a total of six unit cell lengths: at equilibrium as well as values of 4.25, 4.30, 4.35, 4.40 and 4.45~\AA. Density functional pertubation theory\cite{Refson_etal_2006_PhysRevB_DFPTPaper} (DFPT) was used with an $8\times8\times8$ Monkhorst-Pack phonon wave vector grid that was offset to place one of the grid points at $\Gamma$. Convergence tolerance for force constants during the DFPT calculations was set at 10$^{-6}$~eV~\AA$^{-2}$. Long-range electric field effects which lead to LO/TO splitting were accounted for during the calculations and the phonon acoustic sum rule was enforced.

Fourier interpolation was used to generate phonons along high-symmetry directions for the production of dispersion curves, as well as for 6000 wave vectors distributed randomly throughout the Brillouin zone for the production of densities of states.
As shown in Table \ref{ComparePhononsWithOtherCalcs}, $\Gamma$-point frequencies are within a few percent of values obtained through other calculations and experiments. The final phonon dispersion curves shown in Figure \ref{dispersioncurves} are also a close match for dispersion curves obtained previously through DFT and inelastic neutron scattering.\cite{Bohnen_etal_2009_Cu2ONTE,Gupta_etal_2013_ArXiv_AbInitNTEPhonons}

\subsection{Gr\"uneisen parameter calculations}

The Gr\"uneisen parameter relates the vibrational spectrum of a material to its thermal expansion behaviour. At the macroscopic level, the volumetric coefficient of thermal expansion of a material, $\alpha_{\text{V}}$, can be written as\cite{Dove_IntroductiontoLatticeDynamics}
\begin{equation}
\alpha_{\text{V}} = \frac{\overline{\gamma} C_{V}}{B V}
\end{equation}
\noindent where $\overline{\gamma}$ is the macroscopic Gr\"uneisen parameter, $C_{V}$ is the heat capacity at constant volume, $B$ is the bulk modulus and $V$ is the system volume. Since $C_{V}$, $B$ and $V$ are constrained to have positive values, the sign of the thermal expansion coefficient $\alpha_{\text{V}}$ is determined by the sign of $\overline{\gamma}$.

At the microscopic level, the mode Gr\"uneisen parameter, $\gamma_{i,\mathbf{k}}$, is given by\cite{Dove_IntroductiontoLatticeDynamics}
\begin{equation}
\gamma_{i,\mathbf{k}} = -\frac{V}{\omega_{i,\mathbf{k}}}\frac{\partial \omega_{i,\mathbf{k}}}{\partial V}
\end{equation}
\noindent where $V$ is the unit cell volume, $\omega$ the mode frequency, and the indices $i$ and $\mathbf{k}$ refer to an individual phonon by mode and wave vector respectively. $\gamma_{i,\mathbf{k}}$ is normally positive for a given mode, since atomic bonds normally stiffen under compression, thereby increasing phonon frequency. However in some situations, such as for tension-effect modes as described in the Introduction and Figure \ref{tensioneffect}, mode frequency decreases on compression and thus $\gamma_{i,\mathbf{k}}$ is negative.

$\overline{\gamma}$ is calculated by taking the sum of microscopic mode Gr\"uneisen parameter values, $\gamma_{i,\mathbf{k}}$, weighted according to the contribution of each mode to overall heat capacity $C_{V}$.\cite{Dove_IntroductiontoLatticeDynamics} Therefore, phonons with positive $\gamma_{i,\mathbf{k}}$ contribute to PTE and phonons with negative $\gamma_{i,\mathbf{k}}$ contribute to NTE.

Values of $\gamma_{i,\mathbf{k}}$ were calculated for the $a=4.358$~\AA\ equilibrium structure by considering changes in phonon frequency and volume at $a=4.30$~\AA. The result was then converted to a linear colour scale that ranged from red ($\gamma_{i,\mathbf{k}}\le-8$) to white ($\gamma_{i,\mathbf{k}}=0$) through to blue ($\gamma_{i,\mathbf{k}}\ge+8$). 

Plotted dispersion curves were then shaded according to their corresponding value of $\gamma_{i,\mathbf{k}}$ on this scale, whilst bins that made up the plotted density of states were shaded according to the average $\gamma_{i,\mathbf{k}}$ for each bin using the same colour scale. As can be seen in Figure \ref{dispersioncurves}, this process allowed for easy identification and comparison of different modes' contributions to positive and negative thermal expansion.

\subsection{Generation of flexibility models}
\label{GenerationFlexModels}

Simple models were devised to investigate flexibility in the Cu$_2$O framework. Our intention was not to reproduce any other properties of the system other than to abstract the rigid and flexible parts of the framework down to very stiff or complete flexibility.
This approach allowed the simulation of different types of tension effect in this material.

\begin{figure*}[bt]
	\includegraphics[width=\textwidth]{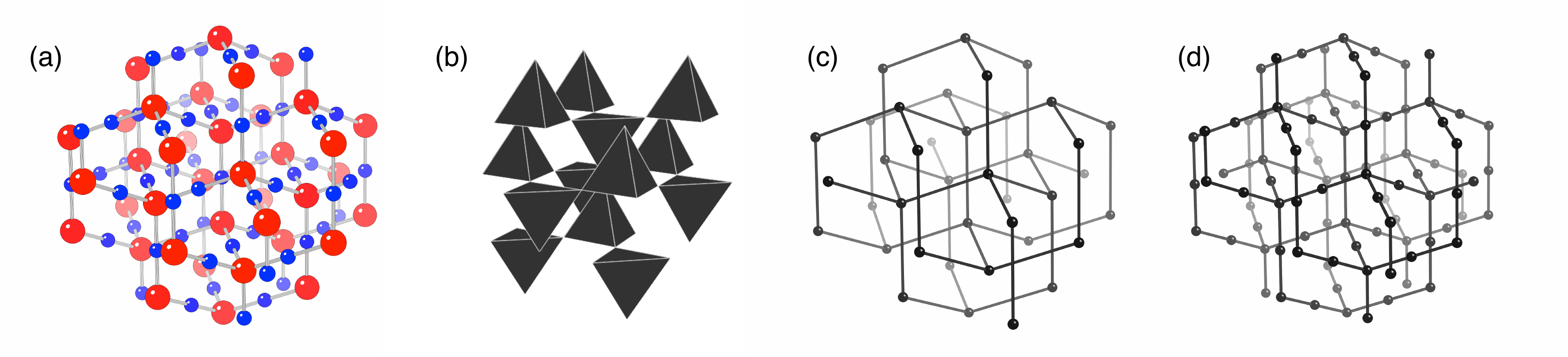}
	\caption{\label{models}(a) The Cu$_2$O structure in ball-and-stick representation: red atoms are O, blue atoms are Cu. (b) Representation of Cu$_2$O Flexibility Model 1, consisting of corner-sharing rigid OCu$_4$ tetrahedra. (c) Representation of Cu$_2$O Flexibility Model 2, consisting of rigid O--Cu--O rods: grey spheres represent flexible linkages. (d) Representation of  Cu$_2$O Flexibility Model 3, consisting of rigid Cu--O rods: grey spheres represent flexible linkages.}
\end{figure*}

Our flexibility models were created in the following manner. The equilibrium Cu$_2$O cell was reconstructed in the program GULP.\cite{Gale_Rohl_2003_MolSim_GULP} A two-body harmonic potential was applied to the Cu--O bond and three-body harmonic potentials were applied to Cu--O--Cu and O--Cu--O bonds. All atoms were assigned zero charge whilst equilibrium bond lengths and angles were chosen such that the cell would remain unchanged upon a zero pressure geometry optimisation.

We were able to control the rigid and flexible regions of the model by changing the force constants in this setup. For this study the force constant for the two-body potential was fixed at 100~eV~\AA$^{-2}$ to model a stiff Cu--O bond. The force constants for the three-body potentials were set to either 100~eV~rad$^{-2}$ or zero, depending on whether the flexibility model in question required those bond angles to be fixed or to vary freely.

Flexibility models considered were:
\begin{enumerate}
\item Rigid OCu${_4}$ tetrahedra wherein Cu--O--Cu bond angles were fixed whilst O--Cu--O bond angles could vary freely.
\item Rigid CuO${_2}$ rods wherein O--Cu--O bond angles were fixed whilst Cu--O--Cu bond angles could vary freely.
\item Rigid CuO rods wherein both O--Cu--O and Cu--O--Cu bond angles could both vary freely.
\end{enumerate}
All of these models are illustrated in Figure \ref{models}. It can be seen in this diagram that Model 1 corresponds to rigid OCu$_4$ tetrahedra; Model 2 to rigid O--Cu--O rods and Model 3 to rigid Cu--O rods. In addition, it should be noted that Model 1 and Model 2 are constrained versions of Model 3. An even further constrained model, where both O--Cu--O and Cu--O--Cu bond angles were fixed, was discounted as this would not have any free deformations and thus could not correspond to any type of tension effect.

Phonons were calculated for all three flexibility models at the same wave vectors as in the \emph{ab-initio} dispersion curves and densities of states. Zero-frequency solutions corresponded to modes involving free deformations of the model structure, whilst solutions with non-zero frequency corresponded to modes that violated the constraints of that model (e.g. Cu--O--Cu bond bending in Model 1).

\subsection{Mapping of flexibility model phonons onto \emph{ab-initio} phonons}

\begin{table*}[tb]
\caption{\label{CompareElasticsWithOtherCalcs}
Cu$_2$O equilibrium elastic and bulk moduli, as obtained through calculation and experiment. In order to account for scatter present in the elastic modulus data as calculated from acoustic modes close to the $\Gamma$ point, values and errors presented here are interpolated from a straight line fit through data points obtained from each volume. Moduli are in units of GPa. Calculated values are close to values obtained via experiment.
The symbol * indicates that, whilst no B$_0$ value was quoted in the original paper, it was inferred here using Equation \ref{BulkModulusFormula}. 
The symbol \dag\ indicates that the experiments were performed at room temperature.
The symbol \ddag\ indicates that the experiments were performed at 4~K.}
\begin{ruledtabular}
\begin{tabular}{l c c c c}
Source & $c_{11}$ & $c_{12}$ & $c_{44}$ & B$_0$ \\
\hline
\hline
This work (PBE+pseudopotentials) & $114.5\pm0.9$ & $87\pm3$ & $7.2\pm0.7$ & $96\pm4$  \\
PBE+pseudopotentials\cite{Cortona_Mebarki_2011_Cu2OHighP} & 116 & 100 & 8 & 106 \\
Inelastic neutron scattering\cite{Beg_Shapiro_1976_PRB_Cu2ONeutronPhonons}\dag & $126.1\pm1.2$ & $108.6\pm0.4$ & $13.6\pm11.4$ & $114.4\pm1.3$* \\
Ultrasonic interferometry\cite{Manghnani_etal_1974_PhysStatSolA_Cu2OElasticConstants}\dag & $122.88\pm0.38$ & $106.50\pm0.71$ & $12.10\pm0.30$ & $111.96\pm0.27$ \\
Pulse echo\cite{Hallberg_Hanson_1970_PhyStatSolB_ElasticConstantsCu2O}\ddag & 121 & 105 & 10.9 & 105.7 \\
\end{tabular}
\end{ruledtabular}
\end{table*}

For each \emph{ab-initio} phonon mode at each wave vector a `match' value, $m_{i,\mathbf{k}}$, was defined as
\begin{equation}
m_{i,\mathbf{k}} = \sum_j\frac{\mathbf{e}_{i,\mathbf{k}}\cdot\mathbf{e}_{j,\mathbf{k}}}{\Omega^2+\omega^2_{j,\mathbf{k}}}
\end{equation}
\noindent where $\mathbf{e}$ is an eigenvector, $\omega$ is a phonon frequency, the index $i$ denotes a mode in the \emph{ab-initio} calculation, the index $j$ denotes a mode in a given flexibility model and the index $\mathbf{k}$ denotes a phonon wave vector. $\Omega$ is an arbitrary scale factor that helps avoid divide-by-zero errors; we set it equal to 1~THz so that a $\omega^2_{j,\mathbf{k}}$ value of 0 gave a $1/(\Omega^2+\omega^2_{j,\mathbf{k}})$ value of 1~THz$^{-2}$.

$m_{i,\mathbf{k}}$ represents the degree to which the flexibility model in question is able to reproduce the mode $i$ at the wave vector $\mathbf{k}$. Scaling by $\Omega^2+\omega^2_{j,\mathbf{k}}$ ensured that only flexibility model modes with zero or close-to-zero frequency contributed to $m_{i,\mathbf{k}}$, as discussed in Section \ref{GenerationFlexModels} above.

Due to the normalisation and orthogonality of the eigenvectors, each $m_{i,\mathbf{k}}$ has a value between 0 and 1. A value of 0 indicates that the flexibility model could not reproduce the \emph{ab-initio} mode eigenvector $i$ at all, whilst a value of 1 indicates a perfect match.

A full set of $m_{i,\mathbf{k}}$ were computed for the equilibrium Cu$_2$O structure. As with the $\gamma_{i,\mathbf{k}}$ plots, $m_{i,\mathbf{k}}$ values were then converted to a linear colour scale. This ranged from white ($m_{i,\mathbf{k}}=0$) through to black ($m_{i,\mathbf{k}}=1$). Plotted \emph{ab-initio} dispersion curves were shaded according to their corresponding value of $m_{i,\mathbf{k}}$ and densities of states were shaded according to the average $m_{i,\mathbf{k}}$ for each bin.

As seen in Figure \ref{dispersioncurves}, this approach yields a convenient visual representation of which real system phonons are reproduced by each flexibility model. This highlights the different types of tension effect-related deformation present in the full phonon spectrum.

\subsection{Elastic modulus calculations}
Acoustic phonon modes with negative $\gamma_{i,\mathbf{k}}$, present in Cu$_2$O,\cite{Bohnen_etal_2009_Cu2ONTE,Gupta_etal_2013_ArXiv_AbInitNTEPhonons} are indicative of elastic softening. Changes in the elastic moduli could therefore be tracked by monitoring the softening of acoustic modes as a function of pressure.\cite{Dove_IntroductiontoLatticeDynamics}

A dynamical matrix solely for the acoustic modes was constructed for all six simulated cells. Monte Carlo minimisation was then used to fit elastic modulus tensor values to both the acoustic dynamical matrix and to the original \emph{ab-initio} acoustic phonon frequencies. The bulk modulus for the equilibrium cell was also calculated using the cubic cell formula\cite{Hill_1952_ProcPhysSocA_ElasticConstantsandModuli}
\begin{equation}
\label{BulkModulusFormula}
B = \frac{c_{11}+2c_{12}}{3}
\end{equation}
\noindent where $B$ is the bulk modulus and $c_{ij}$ are the $i$,$j$\textsuperscript{th} components of the elastic modulus tensor.

Elastic moduli and bulk moduli for the equilibrium cell, compiled in Table \ref{CompareElasticsWithOtherCalcs}, are a good match for other data obtained through theory and experiment.

\section{Results}

\subsection{Thermal expansion in Cu$_2$O}

\begin{figure*}[hbtp]
	\includegraphics[width=\textwidth]{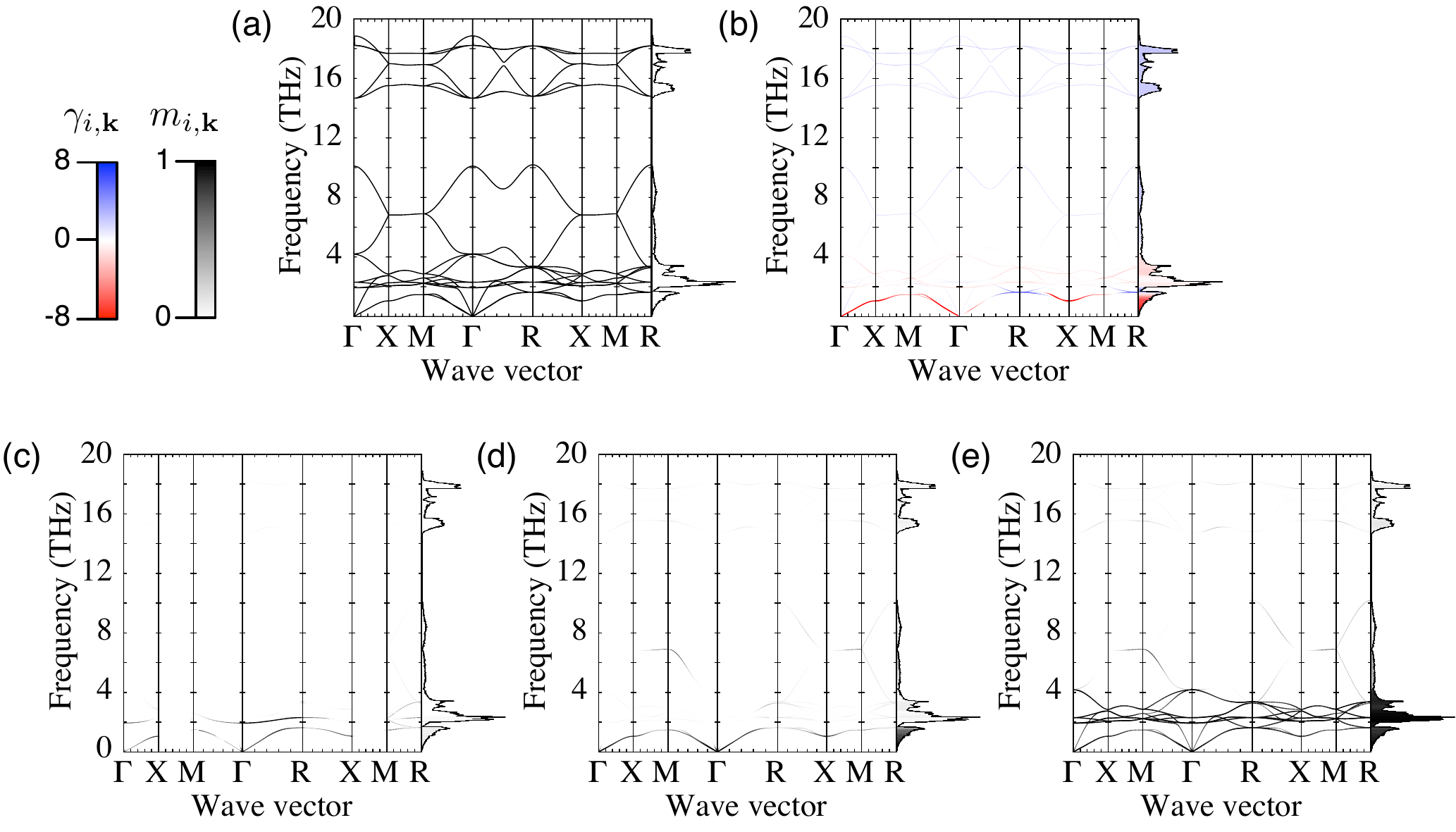}
	\caption{\label{dispersioncurves}(a) Phonon dispersion curves and full Brillouin zone density of states for the equilibrium Cu$_2$O cell. We use the notation of Bradley and Cracknell\cite{Bradley_Cracknell_1972_CrystalBZSymmetryBook} to label special points in reciprocal space: $\Gamma=(0, 0, 0)$, $\text{X}=(0.5, 0, 0)$, $\text{M}=(0.5, 0.5, 0)$, and $\text{R}=(0.5, 0.5, 0.5)$.  Eigenvector matching was employed to differentiate crossings and anti-crossings in the dispersion curves. (b) Shows the same data shaded according to the value of $\gamma_{i,\mathbf{k}}$  of each mode at each wave vector with respect to phonons calculated for a Cu$_2$O cell of $a=4.30$~\AA. The colour scale ranges from red ($\gamma_{i,\mathbf{k}}\le-8$) to white ($\gamma_{i,\mathbf{k}}=0$) through to blue ($\gamma_{i,\mathbf{k}}\ge+8$). Bins that make up the  density of states are shaded according to the average $\gamma_{i,\mathbf{k}}$ for each bin using the same colour scale. (c), (d) and (e) Show the same data shaded according to the value of $m_{i,\mathbf{k}}$ of each mode at each wave vector. The colour scale ranges from white ($m_{i,\mathbf{k}}=0$) through to black ($m_{i,\mathbf{k}}=1$). Bins that make up the densities of states are shaded according to the average value of $m_{i,\mathbf{k}}$ for each bin using the same colour scale. (c) Shows $m_{i,\mathbf{k}}$ values for Flexibility Model 1, (d) Shows $m_{i,\mathbf{k}}$ values for Flexibility Model 2, (e) shows $m_{i,\mathbf{k}}$ values for Flexibility Model 3.}
\end{figure*}

Equilibrium cell phonons, their associated mode Gr\"uneisen parameters and flexibility model match values are all shown in Figure \ref{dispersioncurves}. 
The results are a close match to previous calculations and experimental data.\cite{Bohnen_etal_2009_Cu2ONTE,Reimann_Syassen_1989_PhysRevB_Cu2ORamanPhotoluminescence,Gupta_etal_2013_ArXiv_AbInitNTEPhonons} The largest contribution to NTE in the 0--5~THz range comes from the transverse acoustic modes in the $\Gamma$--X--M region of reciprocal space ($\gamma_{i,\mathbf{k}}\sim-9$ for $\Gamma$ to X and their surroundings, slowly rising towards $-3$ approaching M). There is also some small contribution from several optic modes spanning the Brillouin zone in the 2--5~THz range ($\gamma_{i,\mathbf{k}}\sim-1$). 
The largest contribution to PTE in the 0--5~THz range comes from the three acoustic modes around R ($\gamma_{i,\mathbf{k}}\sim+5$) There is also some very small contribution around $\Gamma$ from the longitudinal acoustic mode and a pair of optic modes ($\gamma_{i,\mathbf{k}}\sim+2$).

Model 1, consisting of rigid OCu$_4$ tetrahedra, is a reasonable match for \textit{some} NTE and PTE modes but not for others, suggesting there is limited correlation between thermal expansion and modes available to this model. The main good match is for a single, nearly dispersionless, weak NTE optic mode at $\sim2$~THz that exists throughout most of the Brillouin zone. This is a moderate match to a RUM consisting of OCu$_4$ tetrahedral rotations and has a particularly strong match in the $\Gamma$--R direction. The same optic mode is also a very weak match for Model 2 (rigid CuO$_2$ rods) but a strong match for Model 3 (which is a less constrained version of Model 1, in that is allows Cu--O--Cu bond bending). We conclude that, away from the $\Gamma$-R region, the mode is forced to undergo some Cu--O--Cu bond bending as a result of constraints of the Cu$_2$O structure and connectivity as well as phonon wave vector---this point is discussed in greater depth in Section \ref{FrameworkFlexNTEDiscussion}. There is therefore some small contribution to NTE from rotations of rigid or near-rigid OCu$_4$ tetrahedra.

Model 2, consisting of rigid O--Cu--O rods, is a strong match to the principal NTE modes highlighted in Figure \ref{dispersioncurves}. The NTE in Cu$_2$O is therefore primarily driven by vibrations of rigid O--Cu--O rods. Weak NTE optic modes that do not match Model 1 are also moderate matches for Model 2. Again, as these modes are also a strong match for Model 3, it follows that the remaining NTE modes are almost a match for Model 2, but are forced to undergo some O--Cu--O bond bending for the same reasons as stated above for Model 1. There is therefore also some small contribution to NTE from motion of near-rigid O--Cu--O rods.

Finally Model 3, consisting of rigid CuO rods, dominates the low energy region of the phonon spectrum, confirming that Cu--O bond stretching is a high energy process. Since the NTE-driving modes described by this model can also be described by Model 1 or Model 2, the extra flexibility afforded to Model 3 (compared the others) does not offer any additional tension effect that leads to NTE in Cu$_2$O.

Putting everything together, we can describe the thermal expansion of Cu$_2$O in terms of its vibrational spectrum as follows: 
the 0--1~THz range is dominated by strong NTE modes in the $\Gamma$-X-M region of the Brillouin zone.
These correspond to vibrations of rigid O--Cu--O rods. 
At $\sim1.8$~THz there is a strong PTE contribution from modes at and around the R point. 
This is expected for structures consisting of a pair of interpenetrating lattices: the R point corresponds to $\Gamma$ for an individual sublattice, and acoustic modes here correspond to the sublattices moving in anti-phase.   

At $\sim2$~THz there is a nearly dispersionless weak NTE mode that spans the entire Brillouin zone and thus forms the large and sharp peak in the low energy phonon density of states.
This mode corresponds to rotations of rigid OCu$_4$ tetrahedra where wave vector allows and to near-rigid OCu$_4$ tetrahedra where constraints of framework connectivity and wave vector mean rigid unit rotations are not possible.
Weak PTE modes around $\Gamma$ also exist in this frequency range; these correspond to rigid Cu--O rod motion and no particular tension effect.

At $\sim3$--4~THz there is a second band of weak NTE modes that also span much of the Brillouin zone, but with a greater dependence of frequency on wave vector than the $\sim2$~THz mode. 
These modes correspond to motions of near-rigid O--Cu--O rods.

Higher frequency modes involve a non-trivial level of Cu--O bond stretching and thus contribute to standard PTE.

\subsection{A predicted phase transition}
\label{Apredictedphasetransition}

\begin{figure}[tb]
	\includegraphics[width=0.4\textwidth]{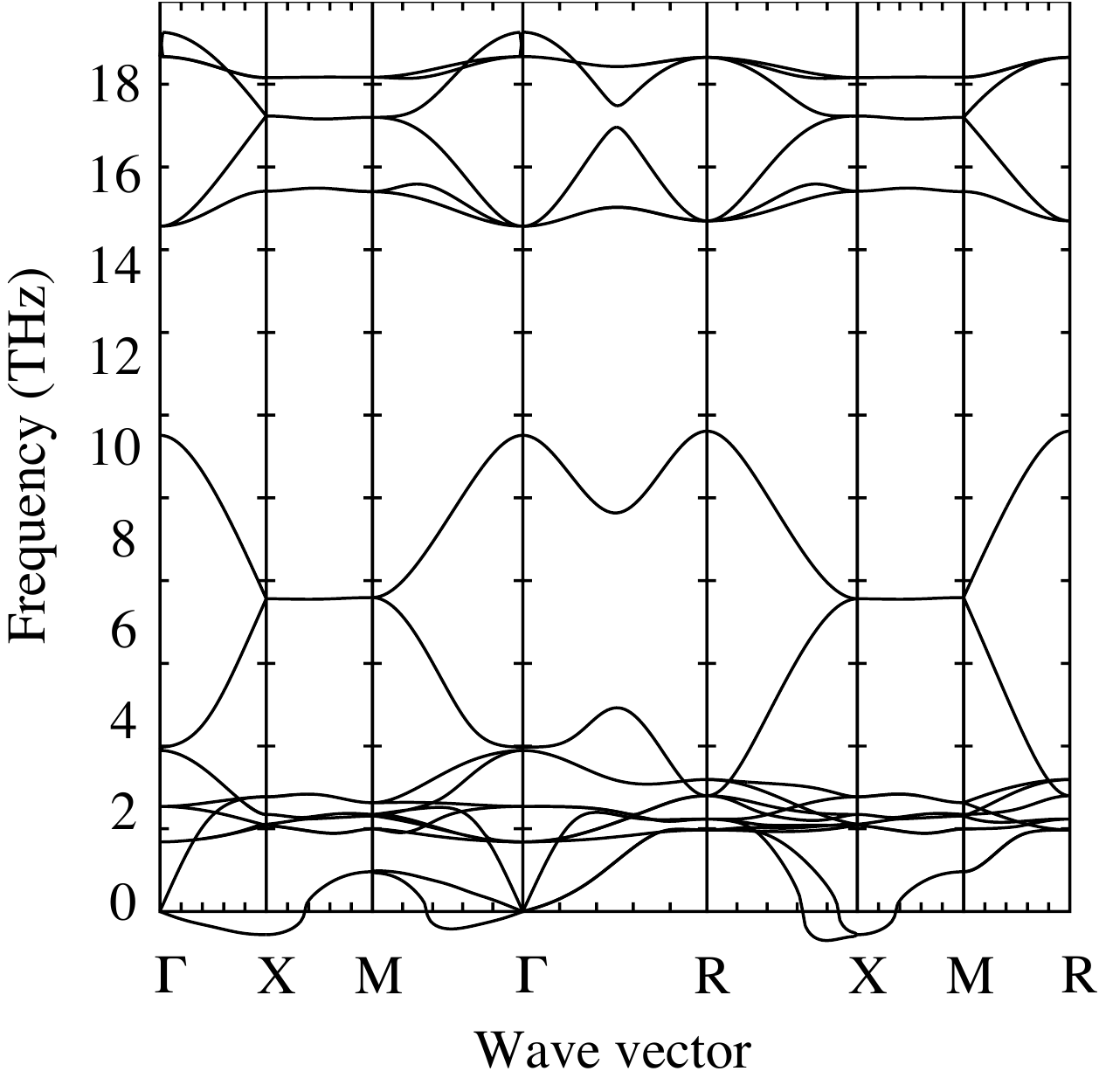}
	\caption{\label{softmodes}Phonon dispersion curves and full Brillouin zone density of states for Cu$_2$O with unit cell parameter $a=4.25$~\AA\, corresponding to an effective pressure of 9.6~GPa. We use the notation of Bradley and Cracknell\cite{Bradley_Cracknell_1972_CrystalBZSymmetryBook} to label special points in reciprocal space: $\Gamma=(0, 0, 0)$, $\text{X}=(0.5, 0, 0)$, $\text{M}=(0.5, 0.5, 0)$, and $\text{R}=(0.5, 0.5, 0.5)$.  Eigenvector matching was employed to differentiate crossings and anti-crossings in the dispersion curves. The $\Gamma$--X acoustic NTE mode is completely soft at this pressure.}
\end{figure}

Softening of the NTE phonon modes as a function of pressure is illustrated in Figure \ref{softmodes}. Zone-centre softening of acoustic branches along $\Gamma$--X corresponds to instabilities in the elastic moduli $c_{44}$ and $c_{11}-c_{12}$ which would, in turn, drive a proper ferroelastic phase transition.\cite{Dove_IntroductiontoLatticeDynamics} 

\begin{figure}[tb]
	\includegraphics[width=0.48\textwidth]{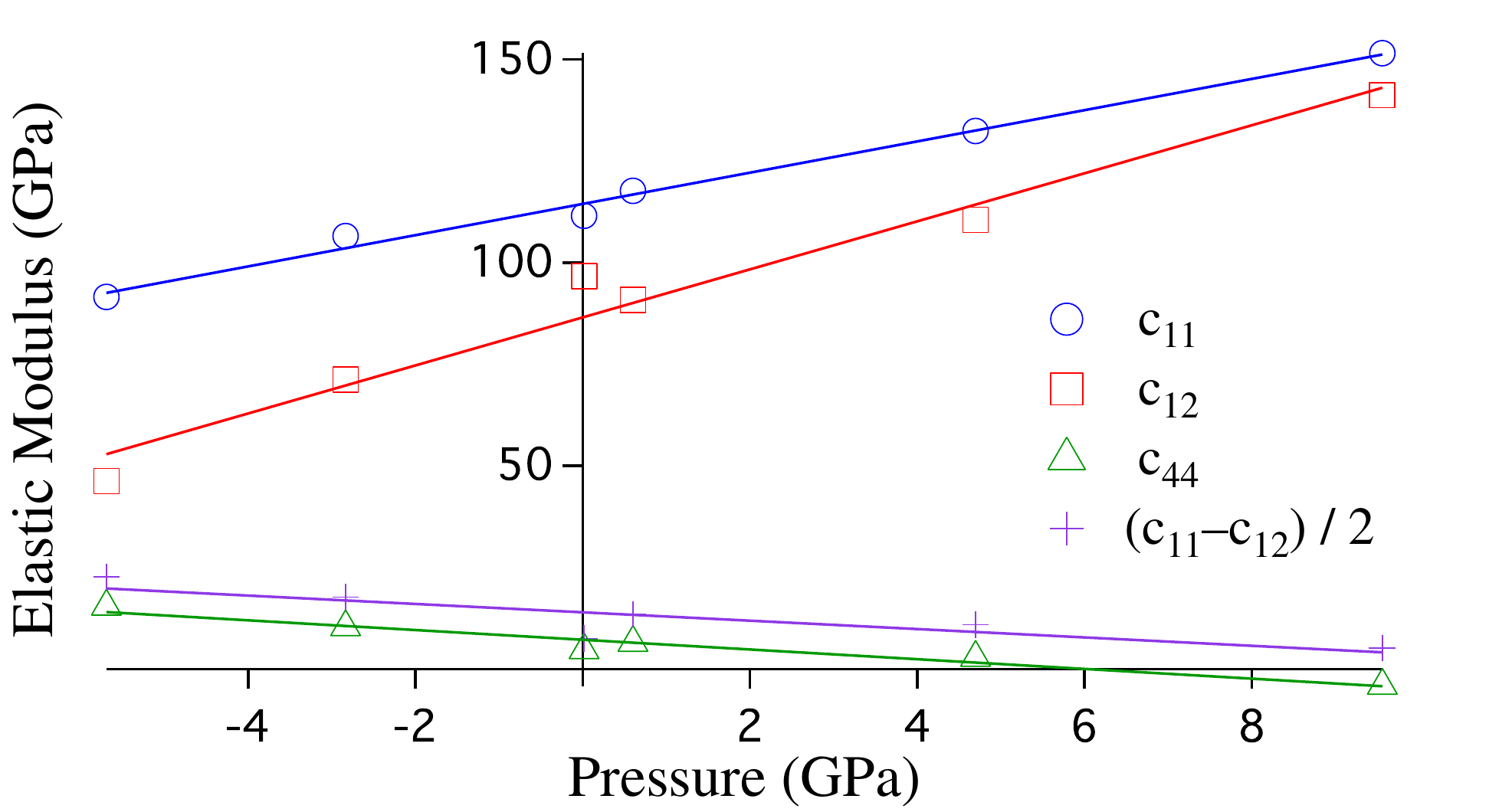}
	\caption{\label{elasticconstantgraph}The elastic moduli of Cu$_2$O as a function of pressure. Straight line fits have been applied to the data for each modulus. $c_{44}$ and $c_{11}-c_{12}$ both soften as a function of pressure. $c_{44}$ reaches zero before $c_{11}-c_{12}$, indicating a ferroelastic phase transition to an orthorhombic structure at $6.0\pm0.8$~GPa.}
\end{figure}

Figure \ref{elasticconstantgraph} shows elastic moduli for all six calculated volumes as a function of effective pressure. The plot shows faster softening in $c_{44}$, indicating a transition to an orthorhombic phase\cite{Dove_IntroductiontoLatticeDynamics} at $6.0\pm0.8$~GPa.

In order to characterise the new phase, another geometry optimisation was carried out on the $a=4.25$~\AA\ unit cell under an applied hydrostatic pressure of 8.5~GPa and using the convergence criteria detailed earlier in Section \ref{Abinitiolatticedynamics}. On this occasion, however, only $P1$ symmetry was applied (though 90$^{\circ}$ cell angles remained fixed). In addition, one cell parameter was changed to 4.26~\AA\ and the fractional coordinate of O at $(0, 0, 0)$ was changed to $(0.02, 0, 0)$. These minor adjustments ensured the geometry optimiser did not become trapped at an energy maximum.

The relaxed cell was confirmed as orthorhombic with space group $Pnnn$. Fractional coordinates of the cell contents remained unchanged from their values in the cubic cell, however the cell parameters at 8.5~GPa became $a=4.256$~\AA, $b=4.261$~\AA\ and $c=4.263$~\AA.

\section{Discussion}

\subsection{Framework flexibility and NTE in Cu$_2$O}
\label{FrameworkFlexNTEDiscussion}

We now understand the driving force behind NTE in Cu$_2$O. It is dominated by low frequency rigid O--Cu--O rod motion, which corresponds to the translational acoustic modes spanning the $\Gamma$--X--M region of reciprocal space. There is some small additional contribution at higher frequencies from rigid and near-rigid OCu$_4$ tetrahedral rotation as well as near-rigid O--Cu--O rod motion, all corresponding to optic modes throughout the Brillouin zone.

The O--Cu--O rigid rod mechanism goes beyond the traditional picture of a tension effect. Instead of transverse vibrations of individual bonds (as illustrated earlier in Figure \ref{tensioneffect}) the mechanism consists of rotations of rigid rods made up of multiple atoms. As these rods rotate they necessarily pull atomic planes together and this motion drives macroscopic NTE.

The NTE in Cu$_2$O can therefore be described as being driven by RUMs, but with the rigid unit recast as an O--Cu--O rod. The additional NTE modes would then be O--Cu--O quasi-RUMs\cite{Hammonds_etal_1996_AmMineral_QRUMPaper} (QRUMs---modes close to being RUMs but which involve some small deformation of a rigid unit due to constraints of framework structure, connectivity and of wave vector), as well as both RUMs and QRUMs of an OCu$_4$ tetrahedral unit.

That NTE in Cu$_2$O is primarily driven by a mode that involves significant deformation of OCu$_4$ tetrahedra had previously been proposed.\cite{Sanson_2009_SolStatSci_TensionLocalDynamicsCuprites,Sanson_etal_2006_PRB_Cu2OAg2ONTE,Chapman_Chupas_2009_ChemMater_NTECuprites} However, this study marks the first time that any proposed mechanisms have been tested against eigenvectors of confirmed NTE modes throughout the Brillouin zone. Furthermore, these previously proposed mechanisms all assumed a Cu--O bond as the basic rigid unit; our work instead shows that an O--Cu--O rigid rod is in fact the fundamental rigid unit as far as NTE behaviour in this material is concerned.

Our analysis of the vibrational spectrum therefore also provides insight into the intrinsic framework flexibility of Cu$_2$O. As known previously,\cite{Sanson_etal_2006_PRB_Cu2OAg2ONTE} Cu--O bond stretching is a high energy process and this is also evident in our analysis, as Cu--O bond stretching modes have been shown to exist only at frequencies greater than 4~THz.

The \emph{lowest} energy deformations (0--1~THz) of the structure are, however, RUMs involving rigid O--Cu--O rods. O--Cu--O bond angle bending does not occur until we reach optic modes at slightly higher frequencies $\ge2$~THz. There is therefore some small, but not trivial, cost to bending the O--Cu--O bond. This is, perhaps, not surprising given the particularly high sensitivity to bond geometry associated with bonds involving d orbitals---as is the case for the Cu centre of the O--Cu--O rod.

Meanwhile, one $\sim2$~THz optic mode consists of either OCu$_4$ RUMs (along $\Gamma$--R) or OCu$_4$ QRUMs (the rest of the Brillouin zone), again as a result of constraints of the framework structure and connectivity as well as wave vector. Cu--O--Cu bond bending has minimal effect on the mode frequency---hence the size of the peak in the phonon density of states at this frequency. There is therefore minimal energy cost to Cu--O--Cu bond bending.

\subsection{Comparison with other NTE materials}

The dominant NTE mode in Cu$_2$O is reminiscent of the mechanism driving weak low-temperature NTE in simple frameworks with a diamond or zincblende lattice such as Si, Ga and CuCl.\cite{White_1993_ContemporaryPhysics_OlderNTEDataReview} NTE in these materials is driven by translational acoustic vibrations exactly like those of the rigid O--Cu--O rod model. This is not surprising given that Model 2 is effectively a pair of interpenetrating diamondoid lattices. Unlike the aforementioned materials however, Cu$_2$O has greater structural flexibility due to the possibility of bending the O--Cu--O bond and thus has additional, albeit weak, NTE mechanisms available to it.

Zn(CN)$_2$, another material composed of two interpenetrating diamondoid lattices, might therefore be expected to behave in a similar fashion to Cu$_2$O. Like Cu$_2$O, NTE in Zn(CN)$_2$ is driven primarily by transverse acoustic modes with some smaller contribution from optic modes.\cite{Fang_etal_2013_PRB_NTEZnCN2} However, whilst the acoustic modes in question are reminiscent of the O--Cu--O acoustic modes in Cu$_2$O (in that they minimise deformation of the Zn--N--C--Zn rod), in practice some minimal deformation of the rod must take place as deformation of the Zn(C/N)$_4$ tetrahedral unit in Zn(CN)$_2$ is a high frequency process. In addition, the dominant NTE-driving acoustic mode in Zn(CN)$_2$ spans the entire Brillouin zone, whilst the dominant acoustic NTE mode in Cu$_2$O only spans the $\Gamma$--X--M region of reciprocal space. This, in turn, means that the NTE in Zn(CN)$_2$ is much larger\cite{Goodwin_Kepert_2005_PRB_ZnCN2NTE} ($\alpha_{\text{linear}}=-16.9$~MK$^{-1}$), 
a difference that can be attributed to the CN bridges in Zn(CN)$_2$ giving its framework more overall degrees of freedom than Cu$_2$O.\cite{Goodwin_2006_PhysRevB_RUMs} 

\subsection{On the use of flexibility model mapping}

The approach taken in this study, whereby phonon eigenvectors generated from simple flexibility models are mapped onto full \emph{ab-initio} calculated modes, has enabled identification of modes in Cu$_2$O that correspond to different types of structural deformation. This, in turn, has allowed for insightful evaluation of proposed NTE mechanisms.

We have found this approach to be more efficient and less error-prone than relying on visual inspection of modes of interest. It has proven useful, in particular, when differentiating Cu$_2$O modes that have very different features but exist at similar energies; such as in the crowded 2--5 THz range of the phonon spectrum.

We anticipate that this same approach will prove highly useful in the analysis of framework flexibility and tension effect-driven NTE in more complex systems such as metal-organic frameworks.\cite{Rao_etal_2008_JPhysCondMat_MOFs} These materials, a number of which exhibit NTE,\cite{Roswell_etal_2005_Science_MOF5,Wu_etal_2008_AngeChem_NTEMOFs} have unit cells typically comprising hundreds of atoms in addition to complex internal bonding. As a result, a full analysis of their dispersion curves using more conventional methods would be a highly complex process.

\subsection{High-pressure phase transitions}

Given that negative Gr\"uneisen parameter indicates mode softening as a function of pressure,\cite{Dove_IntroductiontoLatticeDynamics} one could argue that NTE itself is an outcome of the drive towards a displacive phase transition at pressure. In this case, our results predict a transition to an orthorhombic phase at $6.0\pm0.8$~GPa driven by the strongest NTE modes at $\Gamma$--X. However, in practice, there are numerous conflicting reports of high-pressure phase transitions of Cu$_2$O.

At low temperatures Cu$_2$O is in fact metastable above $\sim4$~GPa,\cite{Sinitsyn_etal_2004_JETPLet_Cu2OHighPAmorphisation,Machon_etal_2003_Cu2OPhaseTransitions} decomposing at equilibrium to form CuO and Cu. Some have predicted a displacive phase transition on the basis of the existence of phonon soft modes \cite{Kalliomaki_etal_1979_PhysStatSolA_Cu2OPhaseTrans} as well as measured\cite{Manghnani_etal_1974_PhysStatSolA_Cu2OElasticConstants} and calculated\cite{Cortona_Mebarki_2011_Cu2OHighP}  softening in $c_{44}$ and $c_{11}-c_{12}$. \emph{Ab-initio} calculations on strained Cu$_2$O cells\cite{Cortona_Mebarki_2011_Cu2OHighP} are a close match to our results and show $c_{44}$ softening fastest, predicting a shear instability at $\sim8$~GPa. A transition was observed experimentally at 5~GPa\cite{Kalliomaki_etal_1979_PhysStatSolA_Cu2OPhaseTrans} using X-ray powder diffraction and a diamond anvil cell. However it was not possible to conclusively identify the new phase at that time.

On the other hand, further X-ray diffraction diamond anvil cell experiments have found no displacive phase transition in the 0--10 GPa range. In one study\cite{Werner_Hochheimer_1982_PRB_Cu2OPhaseTransition} the cubic structure was observed as remaining intact until a known higher-pressure phase transition to a hexagonal structure at 10~GPa. In another study\cite{Sinitsyn_etal_2004_JETPLet_Cu2OHighPAmorphisation} the sample was observed to amorphise at pressure before transforming to the same hexagonal phase at 11 GPa.

Another set of similar experiments did find a displacive transition, but at 0.6\cite{Webb_etal_199_HighPResearch_Cu2OHighP} or 0.7--2.2 GPa,\cite{Machon_etal_2003_Cu2OPhaseTransitions} much smaller pressures than our prediction. This transition was observed to be proper ferroelastic in nature and the new phase was identified as tetragonal\cite{Machon_etal_2003_Cu2OPhaseTransitions} as opposed to our prediction of an orthorhombic phase. Curiously, the structure was also observed to transform into a pseudocubic phase at ~8.5 GPa.\cite{Machon_etal_2003_Cu2OPhaseTransitions}

These inconsistencies can partly be attributed to the fact that phase transitions observed experimentally in Cu$_2$O at high pressure are known to be highly sensitive to the experimental environment, especially the hydrostaticity of the pressure medium.\cite{Kalliomaki_etal_1979_PhysStatSolA_Cu2OPhaseTrans,Machon_etal_2003_Cu2OPhaseTransitions} Furthermore significant peak broadening under pressure obscures any peak splitting,\cite{Machon_etal_2003_Cu2OPhaseTransitions} making pressure-induced displacive transitions in Cu$_2$O difficult to observe experimentally.

Another factor which can complicate the identification of the new phase is the fact that changes in the unit cell for displacive transitions are usually minute. In our predicted orthorhombic phase, the $b$ and $c$ parameters were found to be especially close, within 0.05\% of one another. Such a minute difference may explain why a tetragonal phase is sometimes observed in experiments whilst we predict an orthorhombic structure.

\section{Conclusions}

A full analysis of the NTE mechanism in Cu$_2$O was conducted. This was achieved by mapping different structural flexibility models onto results from \emph{ab-initio} lattice dynamics calculations. The degree to which each model could match the \emph{ab-initio} phonons determined how well it could describe specific modes in the real system. 

It was found that NTE in Cu$_2$O is dominated by low frequency vibrations of rigid O--Cu--O rods. There is also some smaller contribution at higher frequency from rigid and near-rigid OCu$_4$ tetrahedral rotation and near-rigid O--Cu--O rod motion.

It was also found that the primary NTE mode drives a proper ferroelastic phase transition at high pressure. Our simulations predict this to be to an orthorhombic structure with space group $Pnnn$.

\begin{acknowledgments}
LHNR is supported by NERC and CrystalMaker Software Ltd. ALG is supported by EPSRC (EP/G004528/2) and ERC (279705).
Via our membership of the UK's HPC Materials Chemistry Consortium, which is funded by EPSRC (EP/F067496), this work made use of the facilities of HECToR, the UK's national high-performance computing service, which is provided by UoE HPCx Ltd at the University of Edinburgh, Cray Inc and NAG Ltd, and funded by the Office of Science and Technology through EPSRC's High End Computing Programme. 
Additional HECToR calculations were performed via membership of the UK Car-Parrinello consortium, which is funded by EPSRC (EP/F036884/1).
\end{acknowledgments}

\bibliography{2013LHNRCu2OPaper}

\end{document}